\documentclass[12pt]{iopart}
\usepackage{graphicx}

\begin{document}

\title{Discontinuous Jamming Transitions in Soft Materials}

\author{Michael Dennin}

\address{University of California, Irvine\\ Irvine, Ca 92697-4575}
\ead{mdennin@uci.edu}
\begin{abstract}

Many systems in nature exhibit transitions between fluid-like
states and solid-like states, or ``jamming transitions''. Such
transitions can be induced by changing thermodynamic quantities,
such as temperature or pressure (equilibrium transitions) or by
changing applied stress (non-equilibrium transitions). There is a
strong theoretical foundation for understanding equilibrium phase
transitions that involve solidification, or jamming. Other jamming
transitions, such as the glass transition, are less
well-understood. The jamming phase diagram has been proposed to
unify the description of equilibrium phase transitions, the glass
transitions, and other non-equilibrium jamming transitions. As
with equilibrium phase transitions, which can either be first
order (discontinuous in a relevant order parameter) or second
order (continuous), one would expect that generalized jamming
transitions can be continuous or discontinuous. In studies of flow
in complex fluids, there is a wide range of evidence for {\it
discontinuous} transitions, mostly in the context of shear
localization, or shear banding. In this paper, I review the
experimental evidence for discontinuous transitions. I focus on
systems in which there is a discontinuity in the rate of strain
between two, coexisting states: one in which the material is
flowing and the other in which it is solid-like.

\end{abstract}


\maketitle

\section{Introduction}

Most people are familiar with the everyday phenomenon of
``jamming'' as a function of density. This is especially true for
people who drive in heavily populated areas. As the density of
cars on the road increases, the system makes a transition from a
flowing state (traffic moving) to a jammed state (cars at a
stand-still). An interesting feature of jamming is its ubiquitous
occurrence in nature. One observes transitions from freely flowing
behavior to jammed behavior in all types of systems under a wide
range of control parameters. Perhaps the first such transition
that students learn about formally is the freezing of a fluid as a
function of decreasing temperature. Eventually, one also studies
freezing as a function of increasing density. For these familiar
transitions, one is working in the realm of equilibrium
transitions. Significantly less well-understood are jamming
transitions as a function of applied stress \cite{CWBC98,LN98}.
Such transitions are interesting because they are examples of
nonequilibrium transitions. In contrast to equilibrium
transitions, there is no general theory of nonequilibrium jamming
transitions at this time \cite{ITPJamming}. One of the goals of
studies of jamming transitions is to understand how useful the
tools and ideas used for equilibrium systems might be for
understanding the non-equilibrium case.

This concept was expressed by Liu and Nagel when they proposed
that a generalized jamming phase diagram \cite{LN98} was a useful
concept for the characterization of a wide range of phenomena. The
basic idea is that there exists a unified description of jamming
based on considering the behavior of systems as a function of
temperature, the inverse of density, and applied stress (or load).
In this case, systems in the temperature-inverse density plane are
in equilibrium. Additionally, many complex fluid systems are
effectively at zero-temperature because of the large energies
associated with motions of the constituent elements of the
materials. For example, in typical granular materials, the thermal
motion of the individual grains of sand can be ignored. One goal
of the jamming phase diagram is to provide a mapping between
jamming transitions that occur in the different planes. For
example, if one can map the stress axis to the temperature axis,
then elements of our understanding of equilibrium jamming
transitions as a function of temperature can be applied to
non-equilibrium transitions as a function of applied stress.

Before discussing some of the specific models that have been used
in the study of the jamming transition, it is worth defining some
of the terms in more detail. The jamming transition is
fundamentally a transition from a fluid-like state to a solid-like
state (the reverse transition is often referred to as unjamming).
Though this transition is relatively intuitive, it is important to
define each of the states. Because the systems of interest cover
materials ranging from amorphous molecular systems (such as common
window glass) to complex fluids such as foams and granular matter,
a definition based on molecular characteristics is not useful.
Instead, one defines the state of the material based on mechanical
properties. In this case, the two most common properties to
measure are the elastic shear modulus and the viscosity.

Recall that a stress is a force per area applied to the surface of
a material. The shear stress is a force applied parallel to the
surface. A fundamental difference between solids and fluids is the
response to shear stresses. Fluids can not support shear stresses,
so they flow in response to an applied shear stress. Solids can
support shear stresses, so they deform, but do not flow, in
response to an applied shear stress. To describe these different
types of mechanical response, one uses the shear strain, which is
the dimensionless displacement of the material, and the rate of
strain, which is the time derivative of strain. Therefore, one can
define the effective viscosity as the shear stress divided by the
rate of strain and the elastic shear modulus as the shear stress
divided by the strain. Typically, as one approaches the jamming
transition from the fluid phase, the viscosity is found to diverge
(as a larger stress is required to cause the material to flow at
the same rate of strain), and at the transition, the material
develops a non-zero elastic modulus. (For the rest of the paper,
unless specified the elastic modulus will refer to the elastic
shear modulus.) Likewise, if one starts in the solid state and
approaches the unjamming point, the elastic modulus goes to zero
at the transition point, and the system begins to flow. Depending
on the details of measuring the transition, there can be
additional considerations. For a more detailed discussions of the
mechanical response of materials, there are a number of good text
books and references \cite{BOOKS}.

The purpose of this review is to focus on the experimental
evidence for a specific class of jamming transitions:
discontinuous transitions. Therefore, a detailed review of the
theoretical efforts in the area of jamming, and even the
experimental studies of continuous jamming transitions is beyond
the scope of this paper. More complete overviews of the general
jamming transition exist \cite{ITPJamming}. However, it is worth
commenting on some common elements of models of jamming
transitions to set the context for our discussion of the
discontinuous jamming transition.

Most models of jamming focus on the interplay between
heterogeneities in the material and the mechanical response. Three
examples of this will be described briefly. These have been
selected merely to illustrate some of the common elements, and it
is important to realize that there are a broad class of models
that can describe the jamming/unjamming of various materials.

One of the first discussions to use the language of jamming
involved the application of the concept of force chains (the
heterogeneity of interest in this model) in colloidal and granular
systems to the phenomenon of jamming \cite{CWBC98}. The basic idea
is that under an applied load, materials would develop force
chains that could support a specific class of loads elastically.
However, because of the heterogeneous nature of the chains, if a
different load was applied (for example, along a different axis)
the chains would not support the load and the material would
deform plastically. Thus, the material was described as {\it
fragile}.

Another model of materials that experience jamming is {\it soft
glassy rheology} \cite{SLHC97,S98}. Basically, this describes a
large class of materials in terms of local elastic regions with a
distribution of yield stresses (again, a spatial heterogeneity in
the material). (The yield stress is the value of stress at which
an elastic region undergoes plastic deformation or flow.) The
regions are subject to activated dynamics that are controlled by
an effective temperature, which is related to the applied rate of
strain.

The first two models were introduced in the context of complex
fluids, such as colloids, granular matter, and foams. An example
of a model first applied to molecular systems is the concept of
shear-transformation zones (STZ) \cite{FL98}. STZ models have been
used with a high degree of success in describing molecular
plasticity. These are based on the concept of local regions of
bistability in a material. The transition between two different
local configurations of particles provides the fundamental source
of plasticity, and hence, unjamming. In particular, it is
noteworthy that this model is applicable in the limit of no
underlying stochasitc fluctuations and therefore works even at
zero temperature.

Finally, one of the first experimental confirmations that the
jamming phase diagram is a useful concept was reported in
Ref.~\cite{TPCSW01}. This work used three different colloidal
systems to measure jamming transitions as a function of all three
parameters: density, temperature, and applied stress. The
transition was determined by measuring the divergence in the
viscosity of the material or the location at which the elastic
modulus of the material became non-zero. In the context of this
paper, it is important to note the transitions in this case were
all continuous, as opposed to the discontinuous transitions
reviewed in this paper.

\subsection{Continuous versus discontinuous transitions}

In order to distinguish between a continuous and a discontinuous
jamming transition, it is important to understand the two basic
classes of experiments: constant applied stress and constant
applied rate of strain. As discussed, one of the expected control
parameters for the jamming transition is the applied stress.
Therefore, constant stress experiments provide the most direct
probe of the transition. (While the stress across a material can
fluctuate spatially and temporally during flow, by constant
stress, we are referring to the average stress over a large enough
volume to be well defined.) Below a critical stress, the system is
jammed; no flow occurs; and one measures a non-zero elastic
modulus. Above a critical stress, flow is initiated, and one
measures a finite viscosity. The transition can be continuous or
discontinuous, but what does this mean?

\begin{figure}
\includegraphics[width=4in]{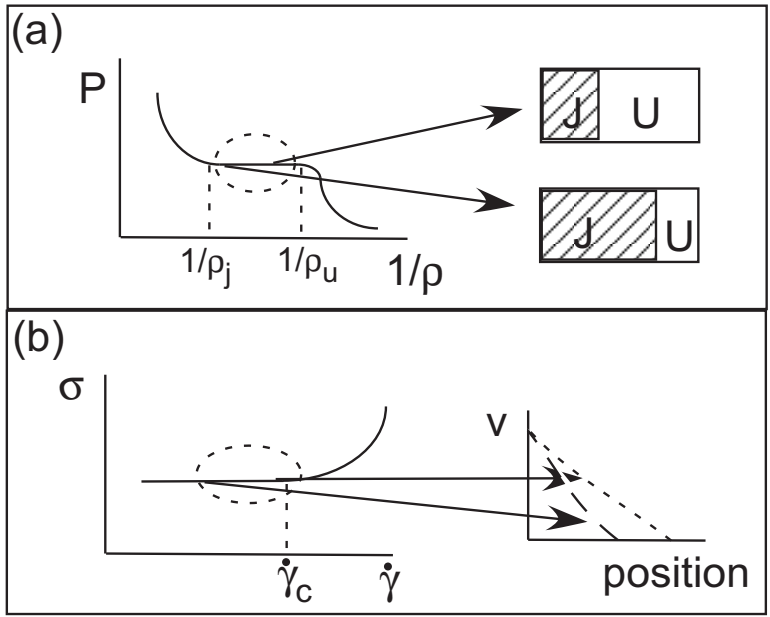}
\caption{This figure provide a schematic illustration of the
analogy between a first order phase transition (a) and a
discontinuous jamming transition (b). In (a), the behavior of the
system as the density is varied is illustrated. In particular, the
coexistence region for which the pressure remains constant is
highlighted. The figures on the right schematically illustrate the
system separating into two regions: one with the jammed (J), or
solid, phase and one with the unjammed (U), or fluid phase. In
(b), the analogous behavior is illustrated for the jamming
transition as a the applied rate of strain ($\dot{\gamma}$) is
varied. In this case, the figures on the right illustrate the
system being divided into two regions: one of flow (non-zero
velocity) and no-flow (zero velocity). In this case, because there
is a critical rate of strain, the slope of the velocity as a
function of position curve is non-zero at the point where the
velocity goes to zero. This indicates the discontinuous nature of
the transition.}\label{phases}
\end{figure}

First, recall the classic first order equilibrium phase transition
with a discontinuity in density - freezing/melting. In this case,
a classic measurement is to fix the pressure and vary the
temperature. At the critical temperature, one measures a
discontinuous change in the density of the material. If one sits
exactly at the transition temperature, one can observe the
coexistence of the two phases. This is often done in the context
of fixing the temperature and varying the volume of the system. In
this case, the pressure of the system is measured. Typically, as
one decreases the volume, the pressure increases until the
transition is reached. At this point, one continues to decrease
the {\it average} density, but the measured pressure stays
constant. Locally, the system phase separates into either the
solid or liquid state, and the average density is a measure of the
relative proportion of one phase to the other (see
Fig.~\ref{phases}).

What is the analog for constant stress or constant rate of strain
experiments as a probe of the jamming transition? In this case,
the stress is analogous to the temperature or pressure, and the
rate of strain is analogous to the density. For experiments in
which the stress is varied, the relevant question is whether or
not the rate of strain increases continuously from zero as the
stress is increased above the critical stress or if there is a
discontinuous jump in the rate of strain. Therefore, in some sense
this is the most straightforward method for probing the continuity
of the jamming transition. For the case of constant rate of strain
experiments, the analogy is to the constant temperature
experiments in which average density is varied. This is a very
common experimental technique for rheological measurements in
either Couette rheometers or cone and plate rheometers.

The basic idea for constant rate of strain experiments is the same
for either Couette or cone and plate rheometers. For the Couette
case, a material is confined between two concentric cylinders. For
the cone and plate, the material is confined between a cone and
plate. Then, one of the confining surfaces is moved at a constant
velocity, applying a constant rate of strain to the system. (Note:
in this case, {\it constant rate of strain} refers strictly to the
dependence in time. Depending on the parameters of the specific
system, the rate of strain may vary in space.) The stress on one
of the walls is then measured - either the stress required to
maintain the constant rotation of the moving wall or the stress
required to hold the other wall fixed - and from the measured
stress/rate of strain relation, the mechanical properties are
determined.

When studying the jamming transition using constant rate of
strain, one can consider both the initial transient response of
the system and the steady-state properties. If the density is
sufficiently low, the system flows immediately, and the stress
immediately jumps to a non-zero value. This corresponds to an
unjammed state. In contrast, if the system is jammed, one observes
an elastic response initially, and the stress increases linearly
with the strain. In this case, the critical stress for unjamming
is eventually reached. At this point, all, or some, of the system
begins to flow. The steady-state behavior depends on the relation
between the average rate of strain and the critical rate of
strain. If the applied rate of strain is sufficiently low, one
observes the coexistence of a flowing and non-flowing region in
the system. This is often referred to as shear-localization,
flow-localization, or shear banding. From the perspective of this
review, an important question is whether or not the rate of strain
is continuous across the boundary from flow to non-flow. A
discontinuity in the rate of strain would be a signature of a
discontinuous jamming transition. As with our analogy to density
in equilibrium transitions, one expects the size of the regions of
flow and non-flow to be set by the average applied rate of strain.
This is illustrated schematically in Fig.~\ref{phases}.

\subsection{Constitutive relations}

Another issue that is worth addressing briefly in the introduction
is the connection between the jamming/ phase transition picture
and ``traditional'' explanations of shear banding for yield stress
fluids. Yield stress fluids are ones that below a critical stress
do not flow, and above a critical stress, they flow. At the most
basic level, this behavior is natural to discuss in the context of
the jamming phase diagram, in which the yield stress corresponds
to the unjamming transition.

Traditionally, yield stress materials are described in terms of
various {\it constitutive} relations. A constitutive relation
typically is given in terms of the stress ($\sigma$) as a function
of the strain ($\gamma$) and/or the rate of strain
($\dot{\gamma}$). This functions is used in Cauchy's equation to
derive the deformation or flow field for the material. A standard
example for yield stress materials is the Herschel-Bulkley model:
$\sigma = \sigma_y + \eta \dot{\gamma}^n$ if $\sigma \geq
\sigma_y$. In this expression, $\sigma_y$ is the yield stress.
Below $\sigma_y$, one can either treat the system as infinitely
rigid (a solid body) or as having a finite elastic modulus and no
viscosity.

How does this differ from the jamming picture? Though in practice,
especially for continuous jamming transitions, one might use
standard rheological relations to describe the mechanical response
of the system, the jamming picture focuses on the transition in
the material between two distinct phases. As we outlined briefly,
many of the models focus on specific structural changes in the
material. When approaching discontinuous transitions, this
provides a useful guide for developing new rheological models.

It is worth commenting on the fact that the concept of yield
stress itself is not completely well-defined, and there exist
various practical definitions of the yield stress. For a good
review of the current understanding of yielding behavior, see
Ref.~\cite{MMB06}. Given this situation, the jamming transition
paradigm, by approaching the question from a different point of
view, might provide insights into our understanding of the
specific case of yield stress materials.

\subsection{Structure of the paper}

There are a number of experimental systems for which discontinuous
transitions have been observed using constant rate of strain. For
some of these, important confirmation of the discontinuity has
been achieved with constant stress experiments. In this review, I
will discuss a number of these and their relationship to each
other. In terms of the jamming transition, the important case is
the coexistence of a flowing state with no flow, and that will be
the focus of this review. But, for comparison, I will also briefly
discuss cases of coexistence of two different flow regimes. The
focus of this review is the experimental results. The systems that
will be discussed are described briefly in Sec.~\ref{exp-sys}, and
the experimental results in Sec.~\ref{exp-res}. However, it is
useful to put these results in the context of proposed mechanisms
for understanding the observed discontinuities in the rate of
strain. This will be done in Sec.~\ref{models}.

\section{Experimental Systems}
\label{exp-sys}

In this paper, I review the results from a number of experimental
systems that exhibit a discontinuous transition from flow to
no-flow. There are two aspects of the work that are important to
consider: (1) the experimental techniques used to study the system
and (2) the details of the system itself. Both of these aspects
play an important role, and it is useful to discuss them
separately. I will first describe the basic experimental
techniques. This will include essential aspects of the method used
to generate the flow, the pros and cons of the methods, and the
methods used to measure the velocity profile (the key measurement
in all of these studies). After reviewing the experimental
techniques, I will discuss the systems that were used for the
studies. In this section, I will discuss the main two-dimensional
system used, bubble rafts, and a number of three-dimensional
materials. This will provide a relatively comprehensive review of
both the various techniques used to probe the discontinuous
jamming transition and the materials for which it is has been
observed.

\subsection{Experimental methods}

\begin{figure}
\includegraphics[width=4in]{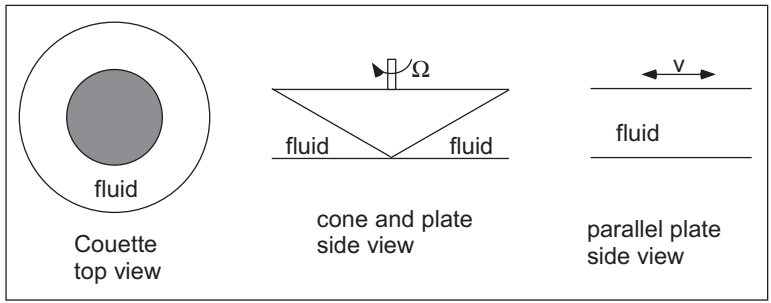}
\caption{Schematic drawings of the three main experimental
geometries. From left to right, Couette viscometer, cone and plate
vsicometer, and parallel plates.}\label{exp-systems}
\end{figure}

There are three standard geometries that have been used to study
shear localization in complex fluids: Couette flow, cone and
plate, and oscillatory planar flow. Figure~\ref{exp-systems}
provides a schematic illustration of each of the geometries. I
will discuss each of these in enough detail to illustrate how the
measurements complement each other.

Couette flow consists of the flow between two concentric
cylinders. This is a standard geometry for making rheological
measurements. The basic measurement in this geometry consists of
measuring the stress as a function of the rotation rate of one of
the two cylinders. An advantage of this system is the fact that it
is extremely well characterized. One knows analytically the
velocity profile in the system for many classes of fluids, even
some that are non-Newtonian. In the cases of interest, the
velocity is purely azimuthal (around the inner cylinder) and only
a function of the radial position between the cylinders. However,
one should be aware that various conventions exist in the
literature for scaling the measured velocity.

In the case of rotating the outer cylinder with rotation rate
$\Omega$, the elastic case corresponds to rigid body motion where
$v_{\theta}(r)=\Omega r$. Therefore, when the outer cylinder is
rotated, the velocities are reported normalized by $\Omega r$,
with $v(r) = v_{\theta}(r)/\Omega r$. In this case, $v(r) = 1$
corresponds to rigid body rotation, which is equivalent to
no-flow. When the inner cylinder is rotated, $v(r) =
v_{\theta}(r)$, and $v(r) = 0$ corresponds to no flow. In some
cases, for rotation of the inner cylinder, the velocity is scaled
by the velocity at the inner cylinder, so that the maximum
velocity is $1$.

An important feature of Couette flow is the fact that there exists
a radial variation in the stress. This is minimized in the case of
narrow gap Couette viscometers, but often plays a role in
measurements with complex fluids. The radial variation of the
stress is a direct consequence of the balance of torques in
steady-state flows and results in a $1/r^2$ dependence of the
stress, independent of whether or not the outer or inner cylinder
is rotated \cite{BAH77}. This fact is important when comparing
different experiments which use Couette flow but rotate different
cylinders. Though the variation of stress plays a role in the
initial observations of the shear localization, if not treated
carefully, it can add a layer of confusion to the interpretation
of the observed shear banding in complex fluids. This is
especially true for the case of a continuous jamming transition.
In this case, it is likely that there would be essentially no
feature in the velocity profile that would distinguish between the
coexistence of two phases and a single constitutive equation for
the material. In contrast, for the work reviewed here, the
drawbacks of the Couette geometry are minimized precisely because
the system exhibits a discontinuous transition. The existence of
the discontinuity in the rate of strain is a clear signal that
would not exist for standarad rheological models.

The other standard rheometer that is used in these studies is a
cone and plate rheometer. One confines the fluid between a plate
and a cone that has a relatively shallow angle. Unlike the Couette
system, this results in a constant stress across the system, to a
very good approximation. Therefore, the cone and plate geometry is
useful for measurements of shear localization that is expected to
depend only on strain rate and not variations in stress. This
geometry is not applicable to the two-dimensional systems.

The final geometry I consider is the flow between two oscillating
parallel plates. As with the cone and plate geometry, this
geometry has a constant stress across the gap. This plays a key
role in unambiguously establishing the coexistence of two
different states of the material, especially at a constant value
of the stress. Another useful feature of an oscillatory geometry
is the ability to consider both the frequency response of the
material and to provide clear tests of the linearity of the
response. When applying a sinusoidal drive to the system, a linear
material will have a sinusoidal response. Additionally, a
perfectly elastic material will have a stress that is in phase
with the strain, and a perfectly viscous material with have a
stress that is $\pi/2$ out of phase. Therefore, if one can measure
the stress in response to an applied strain (or vice versa) one
can simultaneously probe both the viscous and elastic response of
the material.

A final comment on the various techniques used in studying complex
fluids. In all cases, it is critical to understand the role of
wall-slip. Solutions to this problem range from simply measuring
the degree of slip at the wall (for example, \cite{CDSBW06}) to
using various techniques to roughen the walls to ensure that no
slip occurs (for example, \cite{GSD06}).

\subsection{Materials}

The specific systems discussed in this review are all examples of
complex fluids. Here, the term complex fluid is used to refer to
materials that are comprised of the coexistence of two or more
phases or materials, such as foams, colloids, emulsions, slurries,
pastes, and granular materials. Foams are gas bubbles (the
``particle'') with liquid walls, and colloids are solid particles
(the ``particle'') in a fluid \cite{WH99}. A key parameter is the
volume fraction of the ``particle''. For the foam, this is the gas
volume fraction. The limit of a volume fraction of 1 is a
perfectly dry foam. This is an idealization with infinitely thin
walls. For all of these systems, there is a jamming/unjamming
transition as a function of the volume fraction (or density) of
the particles. This is not the focus of this review, but roughly
speaking, as the system increases the density of particles, there
is a jamming transition around random close packing
\cite{ohern_pre03}.

Above this jamming transition, one can refer to the materials as
either in the ``wet'' limit or the ``dry'' limit. I have adopted
this language from the case of foams, for which it refers to the
fraction of fluid in the foam. In the wet limit, the bubbles are
approximately spherical, and there is substantial liquid in the
walls. In the dry limit, the bubbles become polygonal, and the
fluid walls thin. Though based on the description of foam, the
concept of wetness is equally applicable to any complex fluid for
which a volume fraction, or density, of one of the constituents
controls the transition to jamming. Even though there is not a
sharp transition from the wet to the dry limit, we will return to
the concept of wetness in the context of comparing continuous to
discontinuous transitions as a function of applied stress or rate
of strain.

The first specific system I will discuss is the bubble raft
\cite{BL49,PD03,LCD04,D04,GSD06}. A bubble raft consists of a
single layer of bubbles at the air-water interface. In this
regard, it is a quasi-two dimensional system. The bubbles
themselves are still three-dimensional objects. However, the flow
is confined to the plane. Careful optical measurements of the
bubbles in the raft confirm that there is essentially no motion
out of the plane. One also needs to consider the nature of the
interaction between the bubbles and the fluid on which it floats.
In the Couette experiments, this is minimized by flowing the fluid
as well as the bubbles. Since the force between the bubbles and
the water is purely viscous, it is proportional to the difference
in velocity. Therefore, by flowing the fluid with the bubbles,
even though the velocity in the fluid is that of a Newtonian
material, it is still closely matched to the velocity of the
bubbles. Any viscous drag between bubbles and the underlying water
is significantly reduced. In fact, it has been shown that the
bubble-bubble interactions completely dominate over any
interactions between the bubbles and the fluid below them.

All of the experiments reviewed here use bubbles formed by blowing
nitrogen through a solution of 80\% deionized water, 15\%
glycerine, and 5\% miracle bubble solution \cite{LCD04}. The
measurements are all performed using a Couette style system
running under constant rate of strain. In the bubble raft
experiments, the outer cylinder is rotated at a constant rate and
the inner cylinder is held fixed \cite{LCD04}. When comparing
velocity profiles summarized in this paper, it is important to
keep this in mind, as many of the other experiments rotate the
inner cylinder and hold the outer cylinder fixed. The measured
quantity is the azimuthal velocity as a function of the radial
position, $v_{\theta}(r)$, also referred to as the velocity
profile. Because the system is essentially two-dimensional, the
velocity is measured by tracking the motion of individual bubbles
with video images.

A wide range of three dimensional materials (pastes, slurries,
colloids, emulsions, and foam) have been studied. Given the wide
range of materials, I will not detail the specific characteristics
of every system hear. For the three-dimensional materials, a more
interesting feature is the challenge to develop methods of
measuring the velocity profiles, as the materials are
fundamentally opaque. Most of the work discussed in this review
used magnetic resonance imaging (MRI) techniques that are
discussed in detail in Ref.~\cite{RMBBGC02}. Basically, a pulsed
magnetic field method is used to compute the average velocity of
the particles in a small sub-volume of the material. Another
powerful technique is ultrasonic velocimetry \cite{MBC04}.
Briefly, this method uses time-domain cross correlation of
ultrasonic speckle patterns. Typically, this involves including
tracer particles that are used to produce the speckle pattern.
Both the MRI and ultrasonic methods allow for direct measurement
of the average velocity profile inside an opaque materials.
However, in contrast to the bubble raft system, one can not track
the motion of individual bubbles.

Another useful feature of the three-dimensional systems is the
ability to compare studies in the Couette geometry and the cone
and plate geometry. Using both geometries allows for comparison
between homogeneous and inhomogeneous stress. Finally, by using
standard rheometers, it was possible for measurements to be made
at both constant rate of strain and constant stress. The constant
stress experiments are especially important in providing detailed
information about the source of the discontinuity in the rate of
strain that is observed in the constant rate of strain
experiments.

Of the wide range of three-dimensional systems that have exhibited
a discontinuous jamming transition, it is worth briefly commenting
on two systems. One system is a particular colloidal system:
spherical poly-(methyl meth-acrylate) particles with a diameter of
$1.4\ {\rm \mu m}$ in a solvent consisting of a mixture of
cyclohexyl bromide and decalin \cite{CDSBW06}. An important
feature of this system is its optical properties. By appropriate
use of matching index of refractions, the motion of individual
particles can be tracked using confocal microscopy \cite{WCLSW00}.
The other system is an emulsion of castor oil droplets in water
with sodium dodecyl sulfate (SDS) \cite{BMC06}. By varying the
amount of SDS in the water, the authors are able to control the
nature of the interactions between the emulsion droplets. This
provides important insights into the nature of the jamming
transition that will be discussed in later sections.

As discussed in the outline, I have selected to include in this
review a system that exhibits the coexistence between two {\it
flowing} states. Though not an example of jamming per se, it is
very suggestive of a {\it nonequilibrium} transition between two
different fluid phases. Also, the transition is consistent with a
discontinuous transition, providing useful insight into general
discontinuous, nonequilibrium transitions. The system I have
selected to include is worm-like micelles \cite{SCMM03}. It is
representative of behavior in a wide range of materials, including
ordered mesophases and transient gels. These systems are also
notable for the fact that the shear banding is accompanied by
well-defined structural changes in the material. Also, they
provide an example of an additional useful experimental tool for
measuring velocity profiles: dynamic light scattering. The
particular study that I will discuss used a solution of
cetylpyridinium chloride and sodium salicylate in 0.5 M NaCL
brine. The velocity was measured from the interference of a
reference laser beam with the scattered light from a roughly $50\
{\rm \mu m}$ volume of the material. The interference signal from
the two beams exhibits oscillations at a frequency proportional to
the local velocity because of the Doppler shift of the scattered
beam.

\section{Experimental Results}
\label{exp-res}

\subsection{Quasi-Two dimensional System}

\begin{figure}
\includegraphics[width=4in]{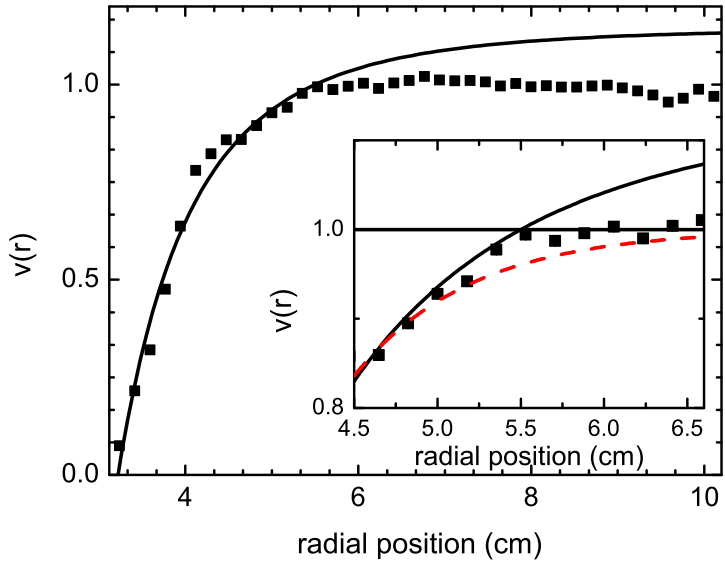}
\caption{The azimuthal velocity as a function of radial position
in a Couette geometry for a bubble raft (solid symbols) (similar
velocity profiles are presented in Ref.~\cite{LCD04} and
\cite{GSD06}). The solid line is the velocity profile in this
geometry assuming that the stress is proportional to the rate of
strain to some power. The insert shows a close up of the
transition from flow to no-flow. The discontinuity in the rate of
strain is indicated by the velocity curve for the power-law fluid.
Also, the case for a fit to an exponential velocity profile is
shown for comparison (red dashed line).}\label{bubblevel}
\end{figure}

For the bubble rafts, the key measurement is the velocity profile,
i.e. the azimuthal velocity as a function of radial position.
Figure~\ref{bubblevel} illustrates a typical velocity profile with
a fit to a the velocity profile that corresponds to a power-law
fluid. (A power-law fluid is one for which the stress is given by
the rate of strain to some power \cite{BAH77}.) It is clear that
the region toward the outer cylinder is undergoing rigid body
rotation ($v(r) = 1$), and that there is a clear transition
between flow (near the inner cylinder) and no-flow near the outer
cylinder.

To determine the nature of the transition, the velocity profiles
were compared to standard yield stress models and to the
coexistence of a power-law fluid and an elastic solid
\cite{GSD06}. Though the discontinuity in rate of strain is
reasonably apparent when the data is fit to a power-law fluid
coexisting with a solid (see insert in Fig.~\ref{bubblevel}),
because of the relatively large discreteness introduced by the
bubbles, a yield-stress model could be forced to fit the data.
However, this model could be ruled out by considering the
resulting parameters from the fit, which were not consistent from
run to run, and in general were unphysical \cite{GSD06}.

An important parameter in these experiments was the critical rate
of strain at the transition point. For sufficiently large system
sizes and rotation rates, there was a single value of the rate of
strain at the transition point \cite{GSD06}. (If the flowing
region becomes sufficiently small, one observes the break down of
continuum behavior in both this system \cite{GSD06} and
three-dimensional foams \cite{RBC05}. However, this behavior is
outside the scope of this review.) The existence of a single value
of the critical rate of strain is consistent with the concept of a
first order phase transition with a jump in the rate of strain
from zero to a critical value.

\begin{figure}
\includegraphics[width=4in]{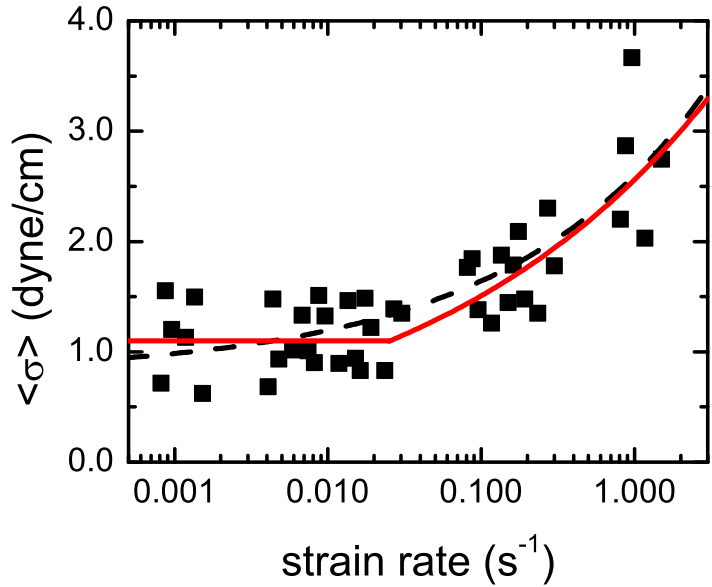}
\caption{Average stress ($<\sigma>$) as a function of the global
rate of strain ($\dot{\gamma}$) for a bubble raft in the Couette
geometry (similar results are given in Ref.~\cite{PD03}). The date
is fit to a Herschel-Bulkley model (dashed black curve) and to a
power-law fluid that has a critical rate of strain (solid red
curve).} \label{bubblestress}
\end{figure}

It is useful to compare this critical value to the behavior of the
stress as a function of the external rate of strain. For a
``phase'' transition, one expects that as the external rate of
strain is lowered, the stress decreases until the critical rate of
strain is reached. At this point, instead of the stress continuing
to decrease, the system will separate into regions of flow and no
flow, where the minimum rate of strain in the flow region is the
critical rate of strain. If this is the case in the bubble raft,
the solid red curve in Fig.~\ref{bubblestress} indicates the
expected stress behavior. A competing view would be that the
material is exhibiting behavior consistent with a yield stress
model (the dashed black line in Fig.~\ref{bubblestress}). As one
can see, the current data is sufficiently noisy that the two
models can not be easily distinguished for the bubble raft, and
future work will be focused on distinguishing these cases.
(Unfortunately, the noise is an intrinsic feature of the bubble
rafts at these very slow rates of strain, and represents a
significant experimental challenge \cite{PD03}).

An interesting feature of the bubble rafts is the ability to
measure fluctuations around the transition. As we progress in our
understanding of the jamming transitions, one expects fluctuations
to play an important role. (Again, this is based on our analogy to
equilibrium transitions.) Also, even in equilibrium transitions,
the question of the exact nature of the interface between two
different phases is an open and very interesting question. For
example, one of the most basic questions is the instantaneous
position of the boundary between flow and no-flow.

\begin{figure}
\includegraphics[width=4in]{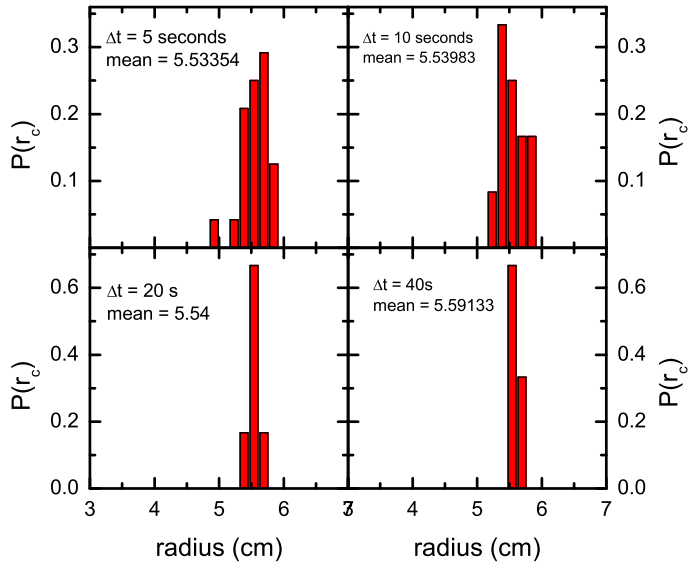}
\caption{Figure 6 from Ref.~\cite{GSD06}. Probability distribution
of measuring a particular value of the critical radius $r_c$ for
the transition from flow to no-flow based on averaging the
velocity over finite time intervals. Four different time intervals
are illustrated for a rotation rate of the outer cylinder of
$0.07\ {\rm s^{-1}}$. The time intervals and mean for each
distribution are indicated in the figure.}\label{criticalrad}
\end{figure}

Because the instantaneous motion of all the bubbles is measured,
one can ask about the location of the transition from flow to
no-flow on short time scales. In this case, ``flow'' is defined on
short time scales by non-affine motion of the bubbles. In other
words, because the outer cylinder is rotating, the elastic motion
of the bubbles is essentially rigid body rotation. Any deviation
from this motion can be interpreted as flow. Therefore, by
considering the displacement of bubbles as a function of position
and measuring the radial position of the initial deviation from
rigid body rotation, the local position of the critical radius can
be determined. From this, one can measure statistics such as the
width of the critical zone and the time required for the averaging
to produce the sharp transition from flow to no flow. As an
initial step, Ref.~\cite{GSD06} reports the distribution of the
measured critical radii as a function of the time used to compute
the ``flow''. The results are reproduced here in
Fig.~\ref{criticalrad}.

\subsection{Three dimensional systems}

\begin{figure}
\includegraphics[width=4in]{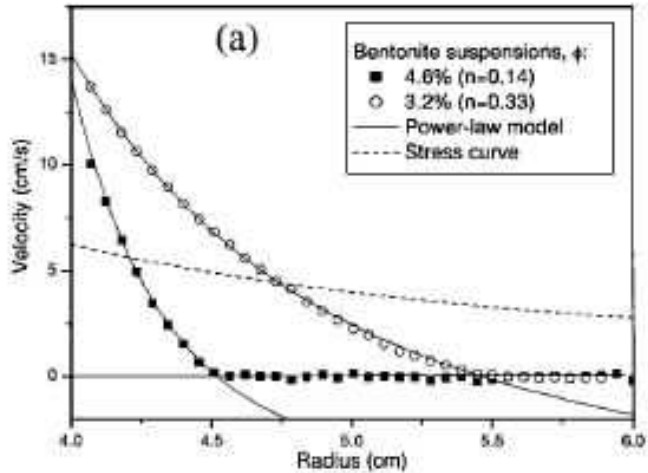}
\caption{An example of a velocity curve obtained using MRI
techniques. This result is taken from
Ref.\cite{CRBMGH02}.}\label{MRIvel}
\end{figure}

As with the bubble rafts, one of the main results for the
three-dimensional studies is the average velocity profile as a
function of the radial position. In Ref.~\cite{CRBMGH02}, a series
of materials are studied using MRI techniques for which the
discontinuity in the rate of strain is clearly demonstrated.
Figure~\ref{MRIvel} is an example of one of the results from
Ref.~\cite{CRBMGH02}. As with the bubble rafts, one can describe
the behavior by fitting the data to the coexistence of a power law
fluid and a solid. This allows for the determination of a critical
rate of strain at the transition point. Again, as long as the
flowing region is large enough, a single critical rate of strain
is determined as a function of external rotation rate.

\begin{figure}
\includegraphics[width=4in]{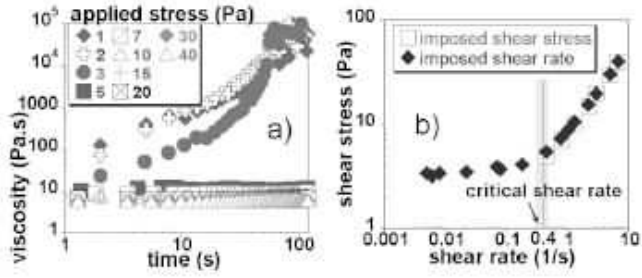}
\caption{Figure 1 from Ref.~\cite{HOBRCD05}. This figure
illustrates both the jump in the rate of strain (as measured by
the viscosity) under conditions of constant stress (left hand
panel) and the behavior of the average stress as a function of
applied rate of strain when a critical rate of strain exists
(right hand panel).}\label{constantstress}
\end{figure}

For the three-dimensional experiments, constant stress experiments
have also been carried out using standard rheometers
\cite{DCBC02,HOBRCD05,MRMB08}. For a traditional yield stress
fluid, one would expect that if the applied stress is below the
yield stress, the material would initial deform some amount (the
elastic response), but it would eventually stop deforming and the
rate of strain would go to zero. As the stress is increased above
the yield stress, the material would flow, with the rate of strain
increasingly continuously from zero. Instead, in the experiments
one clearly observes a jump in the rate of strain from zero to a
non-zero value \cite{DCBC02,HOBRCD05}. This is perhaps the most
straightforward evidence for the discontinuity in the rate of
strain at onset, and an example is given in
Fig.~\ref{constantstress}. It is worth comparing the right-hand
panel of Fig.~\ref{constantstress} to the results for the bubble
raft (Fig.~\ref{bubblestress}). This illustrates the reduction in
noise for the stress measurements in the three-dimensional system.

Additional evidence for the discontinuous jamming transition is
obtained by considering the flow in the cone and plate geometry
\cite{CRBMGH02,MRMB08}. In this case, the stress is now uniform
across the system. Therefore, if the material is exhibiting the
shear banding simply due to a yield stress, shear banding should
not be observed in this geometry. In fact, shear banding is
observed with the MRI experiments, providing additional evidence
for the phase coexistence picture.

\begin{figure}
\includegraphics[width=4in]{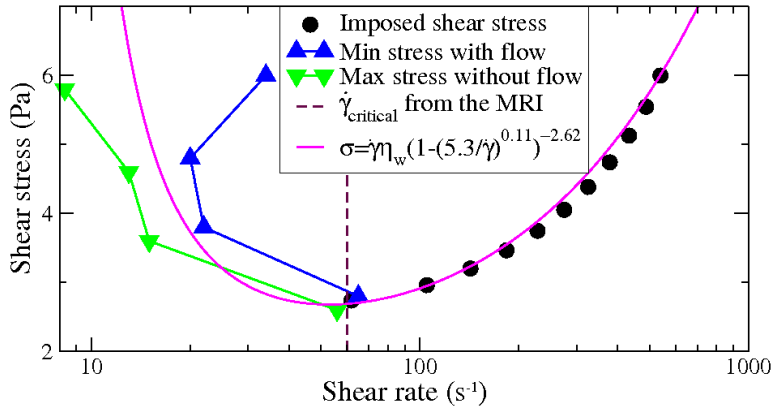}
\caption{Figure~4 from Ref.~\cite{MRMB08}. Illustrates the
measurement of the unstable branch of the stress versus rate of
strain curve. The solid line is the prediction of the model
proposed in Ref.~\cite{MRMB08} to explain the critical rate of
strain. The symbols are data taken under various
conditions.}\label{unstablebranch}
\end{figure}

Finally, as will be discussed in Sec.~\ref{models}, a key element
of many models of shear banding is a region of the stress as a
function of rate of strain curve for which the slope is negative.
This represents a region of unstable flow, so it can not be
observed in steady-state. However, the system can be prepared near
this branch with a fixed applied stress. If the initial rate of
strain is less than the value of the unstable branch, the rate of
strain will continue to decrease. If it is greater than the
unstable branch, it will increase. By locating the transition from
decreasing to increasing rates of strain under a fixed applied
stress, one can determine the location of the unstable branch.
This has been accomplished for at least one colloidal system (see
Fig.~4 from Ref.~\cite{MRMB08}, reproduced here as
Fig.~\ref{unstablebranch}).

An important step in our understanding of the nature of
discontinuous versus continuous transitions is reported in
Ref.~\cite{BMC06}. As described in Sec.~\ref{exp-sys},
measurements were made of the velocity profiles for emulsions
under varying conditions of attraction. The key result was the
observation of a discontinuous jamming transition for adhesive
emulsions and the observation of homogeneous flow for non-adhesive
emulsions. This suggests an important direction for future
theoretical studies of the jamming transition.

\begin{figure}
\includegraphics[width=4in]{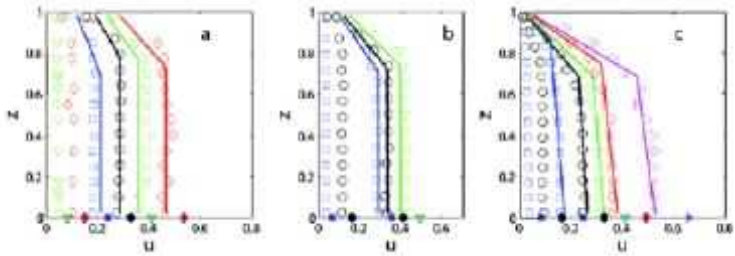}
\caption{An excerpt from Fig.~1 from Ref.~\cite{CDSBW06}.
Illustrated here is the flow profiles for various applied
frequencies as a function of the position (z) in the system. The
profiles are given by the maximal displacement of the particles
(u).}\label{colloids}
\end{figure}

Another important experiment in terms of providing guidance for
the modelling of the discontinuous jamming transition is the work
with the colloidal system described in detail in
Sec.~\ref{exp-sys} \cite{CDSBW06}. In this case, one observes a
critical frequency at which fluid and solid-like behavior coexist.
As with the other experiments, the coexistence point moves as a
function of the applied frequency. An example of the behavior
reported in Ref.~\cite{CDSBW06} is given in Fig.~\ref{colloids}.
The main difference between this system and the other systems is
the fact that the fluid state in this case is Newtonian.

One consequence of the fact that the fluid-like phase is Newtonian
is that this transition is definitely not the result of the
material being a yield stress fluid. There is still the equivalent
of a ``yield stress'' for the system in the sense that there is a
value of the stress which determines when the solid switches to a
fluid. However, the description used to interpret the results
requires a two phase model for the material, but no nonlinearity
is needed in the stress-rate of strain relation.

\subsection{micelles}

\begin{figure}
\includegraphics[width=4in]{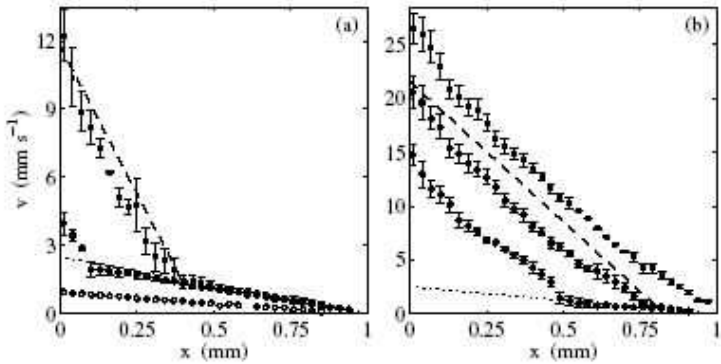}
\caption{Figure~2 from Ref.~\cite{SCMM03} illustrating the
discontinuity in the velocity profile in worm-like
micelles.}\label{micelles}
\end{figure}

A brief discussion of the worm-like micelle system is included as
an important additional example of a shear-band with a {\it
discontinuous} change in the rate of strain. In this case, the
system exhibits behavior that is a classic example of phase
coexistence \cite{SCMM03}. In a standard imposed rate of strain
experiment, the system reaches a stress plateau. At this point,
local velocity measurements establish the existence of two
distinct flow regimes in the material. The two regimes are
characterized by different rates of strain. If the experiments are
carried out as a function of increasing the applied rate of
strain, the region with a higher rate of strain increases in size
until it fills the system. At this point, the measured stress
increases again. Measurements of the material reveal distinct
microstructures in the two flow regimes, consistent with a phase
transition picture. A typical result from Ref.~\cite{SCMM03} is
reproduces here in Fig.~\ref{micelles}. It is worth noting that
shear-banding has also been observed in lamellar phases
\cite{SMC03c}.

The worm-like micelle and lamellar systems are interesting for an
additional reason. As with the bubble raft system, studies of the
dynamics of the interface between the two flow regimes have been
carried out \cite{SBMC03b,LAD06}. As mentioned in the discussion
of bubble rafts, detailed measurements of the dynamics of the
interface at a rate of strain discontinuity promise to provide
additional insight into this phenomenon.

\section{Modelling the rate of strain discontinuity}
\label{models}

The focus of this review is the experimental studies of the
discontinuous jamming transition; therefore, this section is not
intended to be a detailed discussion of the various models that
have been proposed to describe the behavior. Instead, I will
provide a brief overview of some of the essential features of
various models. One motivation for this is the fact that the
experimental evidence for a discontinuous transition in many
different systems is well-established; however, though many
general features of how to model this behavior have emerged, the
theoretical picture at this time is not as well-established.

Once one accepts the general concept of a phase coexistence, one
expects a situation in which there are at least two values of the
rate of strain (zero and the critical rate of strain) for a single
value of the stress. (Recall, that this is depicted schematically
in Fig.~\ref{phases}, and experimental results are illustrated by
Figs.~\ref{bubblestress},\ref{constantstress},\ref{unstablebranch}.)
When this occurs, at fixed rate of strain, the system must ``phase
separate'' to these two different values. The amount of each phase
is determined by a standard application of the Lever rule. Though
a number of models propose more complicated stress versus rate of
strain curves, for example, with regions of stress increasing as
the rate of strain decreases (see for example,
Ref.~\cite{MRMB08}), to explain the data, this simple picture
captures the essence of those explanations.

In the context of this generic concept, it is worth highlighting
one of the very successful models of the discontinuous jamming
transition: the viscosity bifurcation. This concept has been
developed in a series of papers \cite{DCBC02,HOBRCD05,MRMB08}, and
it is useful to point out the most recent application to a
colloidal system \cite{MRMB08}. In this case, a microscopic model
has been proposed based on three elements in the system. First,
aging in the system leads to the development of clusters of
colloidal particles that grow in time. Second, it is the size of
these clusters, and not the particle diameter, that is the
important length scale in determining the rheological response of
the material. And finally, an applied rate of strain breaks up
these clusters. The time scale for the formation and break up of
the clusters are different. Combining these factors results in a
stress as a function of rate of strain illustrated in
Fig.~\ref{unstablebranch}. Particularly interesting is the
experimental confirmation of the region of negative slope
\cite{MRMB08}. Therefore, the model provides a useful microscopic
picture in terms of spatial structures in the material and a
competition of time scales that leads to the discontinuous
transition. Notice, the description in terms of spatial
heterogeneities is similar to that in many earlier models of
jamming. The new feature is the competition of time scales that
produces a critical {\it rate of strain}.

In the context of the experiments reviewed in this paper, it is
worth comparing this detailed model of the viscosity bifurcation
in a particular colloidal system with that proposed as an
explanation for the results of the oscillatory experiments. The
apparent difference is whether or not a nonlinear stress-rate of
strain relation is needed. In the case of the oscillatory
experiments, the data is well-described by the coexistence of a
Newtonian fluid and an elastic solid. However, this is not a
contradiction with the explanation for the viscosity bifurcation.
Instead, one can view it as a special case in which the unstable
branch reduces to a single point, the yield stress. This arises
naturally in the details of the proposed model for how the
colloidal particles cluster \cite{MRMB08}.

Finally, it is worth returning the possible impact of the
``wetness'' of the material and commenting on a different approach
to the question of shear bands that is based on experiments with
very dry foams. In this case, the shear banding is proposed to be
the result of a dynamical focusing of the stress fields
\cite{KD03}. As discussed, the fundamental element of flow in
foams is a T1 event in which bubbles switch neighbors. Modelling
of this process suggests that T1 events can act to focus stress in
such a fashion as to lead to shear banding. In contrast to the
models and experiments already discussed in this paper, this model
produces a continuous transition in that the rate of strain decays
exponentially. However, the decay length can be of the order of a
single bubble, and it has been suggested that this behavior could
account for discontinuous shear bands in which the discontinuity
has not been resolved at the level of single bubbles, or particles
\cite{KSD07}. In considering this proposal, it is worth
reiterating the following aspects of the experiments reviewed in
this paper.

First, even though the measurements of the shear bands using MRI
do not have single particle resolution, the constant stress
measurements strongly support a true discontinuity in the rate of
strain \cite{HOBRCD05}. Second, the two-dimensional bubble raft
measurements \cite{GSD06} and the confocal velocity measurements
in the oscillating shear experiment \cite{CDSBW06} do have a
resolution at the single particle level. Therefore, there is
extremely strong evidence for the discontinuity even at the single
particle level, and it is unlikely that the stress focusing model,
as proposed, can account for the discontinuous shear banding.
However, I include it in this review as this points to an
important direction for future research.

It is not sufficient to explain the source of discontinuous
jamming transitions; one also wants to understand why some
transitions are continuous and others are discontinuous. A common
feature of the experimental systems reviewed in this paper is that
they can be considered to be in the relatively ``wet'' limit.
Again, there is no precise definition of this at the current time.
But, the dynamical model based on T1 events is definitely
applicable to a dry foam, in which the dynamics of the films, and
not the bubbles, dominates \cite{WH99}. Therefore, an important
question for future research is developing a workable definition
of wetness and determining if this acts as a control parameter for
the transition from continuous to discontinuous transitions.
Returning to the analogy with equilibrium transitions, as with the
gas-liquid transition, one might reasonably expect there to exist
the equivalent of a well-defined critical point at which the
transition switches from continuous to discontinuous. Another
important direction to consider is the connection, if any, between
``wetness'' and the results with adhesive emulsions \cite{BMC06}.

\section{Summary}

I have reviewed a diverse set of experiments that demonstrate
strong evidence for the existence of a discontinuous jamming phase
transition. One consequence of such a transition is the existence
of shear bands with a discontinuity in the rate of strain. In the
case of a constant applied rate of strain, this results in
behavior reminiscent of a classic first order phase transition in
which the measured stress is constant for a range of applied rates
of strain. In this range, the system divides into two distinct
regimes with different rates of strain, which on average,
correspond to the fixed applied rate of strain. An exciting
feature of this behavior is the relatively wide range of systems
in which it is observed, including both three-dimensional and
quasi-two dimensional systems.

A generic picture is presented for understanding the existence of
discontinuous jamming transitions: the existence of an unstable
branch in the stress-rate of strain relation. As an example, a
specific model for a viscosity bifurcation based on the
competition between the applied shear and aging in the system is
reviewed. In this case, the shear breaks up clusters of the
material as the aging increases cluster sizes. This competition
results in a model of the stress as a function of rate of strain
that has a critical rate of strain. Essentially, the existence of
a critical rate of strain can be understood in terms of the
competition between rates for the two different processes in the
system. Having this physical picture for one system, it is
interesting to ask about the application to other very different
systems, such as the bubble rafts. In the bubble rafts, there has
been no observation of aging at this point, but this does not rule
out other similar competition of time scales. Despite the
existence of a number of promising models, there are many open
theoretical questions. For example, the issue of predicting
whether or not a particular jamming transition is continuous or
discontinuous is raised.

In conclusion, we are at an exciting point in the study of the
jamming transition. There is sufficient experimental evidence to
take seriously the existence of a discontinuous jamming
transition. However, there are still a number of open questions
that require both additional theoretical and experimental efforts.

\section{Acknowledgements}

There are many people who have provided insights and exciting
discussions of the jamming transition. I would like to thank A.
Liu and D. Durian for introducing me to this exciting field and
all of their discussions. I would like to thank P. Coussot and D.
Bonn for their discussions of the three-dimensional systems and
the connections with the bubble raft. I would like to thank K.
Krishan for his discussions of the manuscript and jamming in
general. Finally, I would like to thank all the authors that were
willing to allow use of their figures in this review.

\section{References}

\bibliographystyle{unsrt}

\end{document}